\documentclass[12pt]{article}
\usepackage{epsfig}
\usepackage{psfrag}
\usepackage{enumitem}
\usepackage{latexsym}
\usepackage{indentfirst}
\usepackage{fancyhdr}
\usepackage{amssymb}
\usepackage{amsmath}
\usepackage{amsfonts}
\usepackage{pifont}
\usepackage{cite}
\usepackage{bbold}
\usepackage{color}
\usepackage{graphics}
\usepackage[center,footnotesize,hang]{subfigure}
\usepackage{url}
\usepackage{array}

\def\be{\begin{equation}}
\def\ee{\end{equation}}
\def\bc{\begin{center}}
\def\ec{\end{center}}
\def\bea{\begin{eqnarray}}
\def\eea{\end{eqnarray}}

\newcommand{\ba}{\begin{array}{c}}
\newcommand{\bad}{\begin{array}{ccc}}
\newcommand{\ea}{\end{array}}

\newcommand{\nova}{NO$\nu$A~}
\def\nn{\nonumber}

\begin{document}

\begin{titlepage}
\hfill{RM3-TH/13-8}
  \vskip 2.5cm
   \begin{center}
    {\Large\bf  Checking Flavour Models at Neutrino Facilities}
   \end{center}
  \vskip 0.2  cm
   \vskip 0.5  cm
  \begin{center}
\vskip .2cm
{\large Davide Meloni}~\footnote{e-mail address: meloni@fis.uniroma3.it}
\\
\vskip .1cm
Dipartimento di Matematica e Fisica, Universit\`a di Roma Tre 
\\ 
INFN, Sezione di Roma Tre, \\
Via della Vasca Navale 84, I-00146 Rome, Italy
\end{center}
\vskip 0.7cm

\begin{abstract}
In the recent years, the industry of model building has been the subject of the intense activity, especially 
after the measurement of a relatively large values of the reactor angle. 
Special attention has been devoted to the use of non-abelian discrete symmetries, thanks to their ability of reproducing 
some of the relevant features of the neutrino mixing matrix. In this paper, we consider two special relations between the leptonic mixing 
angles, arising from models based on $S_4$ and $A_4$, and study whether, and to which extent,  they can be distinguished at superbeam facilities,
namely T2K, NO$\nu$A and T2HK. 
 
\end{abstract}
\end{titlepage}

\section{Introduction}  
The recent measurement of a non-vanishing $\theta_{13}$ by Daya Bay \cite{An:2012eh} and RENO
\cite{Ahn:2012nd} has  exerted some pressure on models for neutrino mixing  based on the permutation groups (like $A_{4}$ and $S_4$, 
\cite{Altarelli:2010gt}), 
as they are generically constructed to give at leading order very specific patterns in which $\theta_{13}=0$ and the other 
angles are also completely fixed.
Corrections from the charged sector or next-to-leading contributions to the neutrino mass matrix have to be
invoked to correct such patterns and make the models compatible with the experimental data.
The usual approach to model building is that of considering a Lagrangian invariant under a flavour group $G$ and to subsequently break 
$G$ into two different subgroups in the charged lepton and neutrino sector, is such a way to create two different rotations, responsible
for a non-diagonal Pontecorvo-Maki-Nakagawa-Sakata ($U_{PMNS}$) mixing matrix. The structure of $G$ can also be reconstructed from the residual symmetries of the mass matrices after symmetry breaking;
for example, using the criterion that a flavour group should be
obtained from the neutrino mixing matrix  without parameter tuning, it was shown in \cite{Lam:2008sh} that
the minimal group containing all the symmetries of the neutrino mass matrix and leading to 
the tri-bimaximal mixing (TBM \cite{TBM}) is $S_4$. The fact that the mixing angles are fixed to well defined values is the consequence of forcing all 
the symmetries of the mass matrix to belong to $G$. Moving from this consideration, in \cite{Hernandez:2012ra} a different point of view was 
adopted: they assumed that the residual symmetries in both the charged lepton and neutrino 
sectors are one-generator  groups. Indicating with $S_{i}$ and $T_{\alpha}$ ($\alpha=e,\mu,\tau$) the generators of the $Z_2$ and $Z_m$ discrete symmetries of the 
neutrino and charged leptons mass matrices, the previous condition implies that $\{S_{i},\,T_{\alpha}\}$  form  a set 
of generators for the flavor  group $G$ for given $i$ and $\alpha$,
with the meaning that all other symmetries appear accidentally. The structure of the generators is restricted by the additional 
requirements to be  elements of $SU(3)$, for which $Det[S_i] =Det[T_\alpha] = 1$, so they can be written as:
\bea
S_1 &=&{\rm diag}(1,-1,-1)  \,, \quad \nn
S_2 ={\rm diag}(-1,1,-1)   \,, \quad
S_3 ={\rm diag}(-1,-1,1)  \\
T_e & =&{\rm diag}(1, e^{2\pi i k/m}, e^{-2\pi i k/m}) \,, \quad 
T_\mu  =  {\rm diag}(e^{2\pi i k/m}, 1, e^{-2\pi i k/m} ) \,, \\
T_\tau  &=&  {\rm diag}( e^{2\pi i k/m}, e^{-2\pi i k/m}, 1 ) \,. \nn
\eea
The definition of $G$ requires a relation linking  $S_{i}$ and $T_{\alpha}$, assumed to be \\
$
(S_{i}T_{\alpha })^p = ( U_{PMNS}S_iU_{PMNS}^\dagger T_\alpha)^p = I \,.
$
The lack of additional symmetry in $G$ has the direct consequence that 
the mixing angles are not all fixed (like in the TBM) but rather present some interesting correlations, or sum rules, that 
open the possibility to reconcile the 
predictions of the permutation groups with the experimental data already at leading order (see also \cite{Feruglio:2012cw} 
for similar sum rules obtained in the context of $S_4$ and \cite{Ge:2011qn} for sum-rules from residual $Z_2$ simmetries).
The question we want to analyze in this paper is whether such correlations can be tested at neutrino facilities or, in other words, if model
comparison and selection can be achieved at currently taking data or planned superbeams. 
It is clear that if two models live in completely 
different regions of the parameter space (given by the spanned values of all $\theta_{ij}$ and the leptonic CP phase) the measurement of the mixing 
parameters with huge precision will give the answer; however, we are still away from such an idealized situation, at least for what concerns the 
CP phase, and it is necessary to evaluate the performance of the neutrino facilities to face this problem. In this respect, we have selected 
two models from \cite{Hernandez:2012ra}, called $1T$ and $2T$, which have been shown to be compatible with the current experimental data in the 
neutrino sector and with the hypothesis of TBM, and have used their different correlations  to compute and compare (in a $\chi^2$ analysis)
the expected event rates at T2K, NO$\nu$A and T2HK, with the aim of identifying the regions in the $(\theta_{13},\delta)$-plane 
where the models can be distinguished at some confidence level.
An interesting  work along similar lines has been recently presented in \cite{Ballett:2013wya}, where the main focus 
was on the ability of next-generation of neutrino oscillation experiments to constraints correlations involving $\theta_{23}$, $\theta_{13}$ and $\cos \delta$.
We differ from \cite{Ballett:2013wya} in that we consider different neutrino facilities, we use non-linear relations between the 
oscillation parameters and adopt a different statistical analysis with the purpose, given the lack of information on the CP phase, 
to present exclusion regions directly in the  $(\theta_{13},\delta)$ parameter space. 
It is important to stress again that such correlations are {\it leading order} predictions, in the 
sense that they are derived from group theoretical considerations and do not take into account possible higher order effects into the 
lepton mass matrices of new-physics effects \cite{arXiv:0809.1041}, otherwise model-dependent features will appear with 
the main effect to spoil the sum rules and introduce additional indetermination of the parameter spaces where the models live. We do not 
take into account this possibility, as we are mainly interested to check whether the easiest case (validity of the sum rules) can be addressed 
at neutrino experiments.
We revise the useful neutrino transition probabilities in Sect.\ref{sec2}, where we also introduce the models $1T$ and $2T$ and discuss the 
parameter spaces allowed by the correlations; in Sect.\ref{sec3} we introduce the neutrino facilities used in our numerical simulation and discuss the 
results of the statistical analysis performed to distinguish the models. Our conclusions are drawn in Sect.\ref{conc}.

\section{Setting the background}
\label{sec2}
\subsection{The relevant transition probabilities}
Since we are interested in the performance of superbeam facilities, it is enough to consider the 
$\nu_\mu \to \nu_e$ appearance and $\nu_\mu \to \nu_\mu$ disappearance probabilities (and their CP-conjugate).
Given the relatively large $\theta_{13}$, we consider the probabilities up to first order in the small 
parameter $r=\Delta m^2_{sol}/\Delta m^2_{atm}\sim 0.03$ \cite{Akhmedov:2004ny} while keeping their exact dependence on $\theta_{13}$. 
In vacuum they read:
\bea
P_{\mu e}&=& \sin^2 2\theta_{13} s^2_{23} \sin^2
\Delta  - r\,\left[\Delta s^2_{12} \sin^2 2 \theta_{13}
s^2_{23} \sin 2 \Delta\right. \nonumber \\
&&\left.+ \Delta \sin 2 \theta_{12} s_{13} c^2_{13} 
\sin 2 \theta_{23} (- 2 \sin \delta_{\mathrm{CP}} \sin^2
\Delta + \cos \delta_{\mathrm{CP}} \sin 2 \Delta )\right] \,,\\ && \nonumber\\
P_{\mu\mu}&=&1-\sin ^2\Delta \left[c_{13}^4 \sin ^2(2  \theta_{23})+s_{23}^2 \sin^2(2 \theta_{13})\right] + 
r \nonumber \\
&&\,\left\{
\Delta  \sin 2\Delta\left(c_{13}^2 \left(\sin^2(2\theta_{23}) \left(c_{12}^2-s_{12}^2
s_{13}^2\right)-4 s_{13} \cos\delta \sin 2\theta_{12}) \sin^3(\theta_{23}) 
\cos(\theta_{23})\right)\right. \right.\nonumber \\
&&\left.\left.+s_{12}^2 s_{23}^2 \sin^2(2 \theta_{13})\right)\right\}\label{eq:Pmumu}
\,,
\eea
where $\Delta \equiv \frac{\Delta m^2_{31} L}{4 E_\nu}$, $s_{ij} =\sin \theta_{ij}$ and 
$c_{ij}=\cos \theta_{ij}$.
Notice that: 
\bea
\label{eq:Panti}
P_{\bar{\alpha}\bar{\beta}} &=& P_{\alpha\beta}(\delta_{\rm CP} \to -
\delta_{\rm CP}) \\
P_{\beta\alpha} &=& 
P_{\alpha\beta}(\delta_{\rm CP} \to - \delta_{\rm CP})\,, \qquad \alpha,\beta = e,\mu,\tau \,.
\eea
As it is well known, $P_{\mu e}$ is mainly dependent of $\theta_{13}$ and $\delta$ whereas $P_{\mu \mu}$ is recognized to be 
more sensitive to the atmospheric parameters; although the dependence on $\delta$ is suppressed by the small $r$, 
the approximation adopted shows that $\theta_{13}$ already appears at leading order.
We then expect that flavour models with different parameter spaces, that is with the mixing parameters living in different regions, 
are also characterized by different  transition probabilities that, extracted from the experimental data, 
can help in distinguishing among them. In our numerical computations we consider the mixing angles to vary within the 2$\sigma$ intervals 
taken from \cite{Fogli:2012ua}:
\bea
\sin^2 \theta_{23}&=& 0.386^{+0.062}_{-0.038}   \nn \\
\sin^2 \theta_{13}&=& 0.0241^{+0.0049}_{-0.0048}  \label{bounds} \\
\sin^2 \theta_{12}&=& 0.307^{+0.035}_{-0.032}\nn\,,
\eea
whereas the CP phase is left free to vary in the whole $[0,2\pi)$ range. We consider the mass differences as constant 
quantities,
$\Delta m^2_{31} = 2.4 \times 10^{-3}$ eV$^2$, $\Delta m^2_{21} = 7.5 \times 10^{-5}$ eV$^2$, since the models 
studied in this paper do not give any information on the neutrino masses.

\subsection{A summary of the models $1T$ and $2T$}
In this section we recall the main features of the correlations arising from two different models 
discussed in  \cite{Hernandez:2012ra}, of which we also follow the nomenclature. Both of them have 
$T_\alpha = T_e$.
The first model, called $1T$,  uses the generator $S_1={\rm Diag(1,-1,-1)}$ and the pair of values
$(p,m)=(4,3)$, which corresponds to the group $S_4$.
The obtained relations among the mixing angles are:
\be
\cos^2\theta_{12}  = \frac{2}{3 \cos^2\theta_{13}} \,.  
\label{S4a}
\ee
and
\be
\tan2\theta_{23} = -  
\frac{ 1 - 5 s^2_{13}}{2\cos\delta s_{13} \sqrt{2(1 -  3 s^2_{13})}} \,,
\label{th23zzz}
\ee
also obtained in the explicit model of Ref.\cite{Lin:2009bw} 
and further studied in \cite{Altarelli:2012bn}.

\begin{figure}[h!]
\begin{center}
\includegraphics[scale=.5]{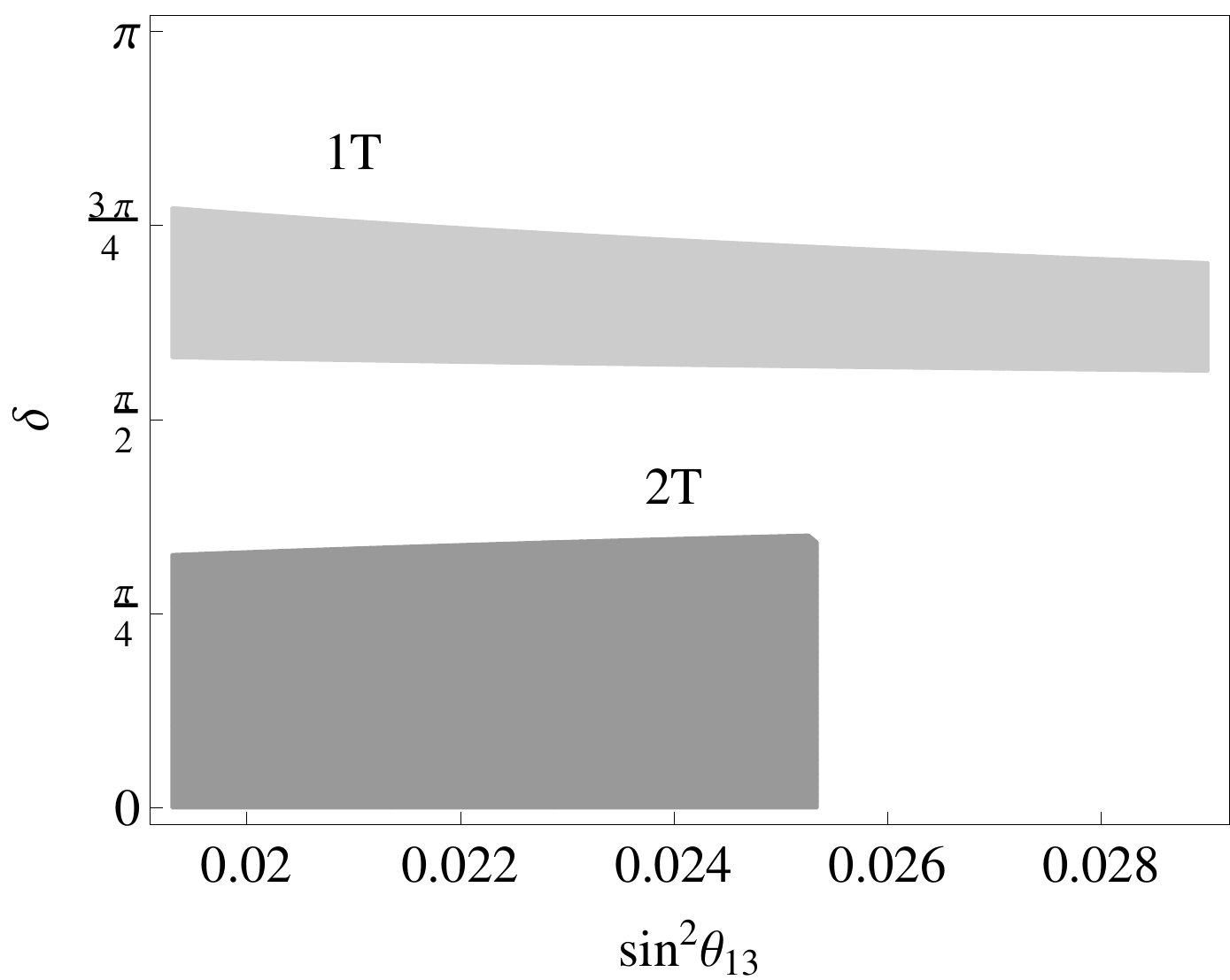} 
\caption{\label{figt1limits}\it Allowed values of $\delta$ as a function of $\sin^2 \theta_{13}$ as derived imposing the correlations 
among the mixing parameters, eqs.(\ref{S4a})-(\ref{th23zzz}) for the model $1T$ and  eqs.(\ref{A4})-(\ref{eqqq}) for $2T$.}
\end{center}
\end{figure}

For any values of $\theta_{13}$, the first relation always gives an acceptable value of the solar angle, within the 2$\sigma$ bounds quoted in 
 eq.(\ref{bounds}), so this relation does not set any restriction on the reactor angle. It has to be noted, however, that 
the dependence on the cosinus function forces $\sin^2\theta_{12}$ to be around $0.31-0.32$, very close to the current central value.
On the other hand, eq.(\ref{th23zzz}) imposes a constraint on the possible pairs of $(\theta_{13},\delta)$ needed to fulfill the bounds
for $\theta_{23}$ in eq.(\ref{bounds}); in particular, the value of the CP phase can never be maximal in this model and, in order to 
have an atmospheric angle in the first octant, the relation $\cos \delta > \pi/2$ must hold.
The exact bounds in the $(\theta_{13},\delta)$-plane can be derived numerically from eq.(\ref{th23zzz}) and are shown 
in Fig.(\ref{figt1limits}) where, as expected, no restriction on the reactor angle is present and 
the CP phase is limited in a horizontal band above maximal violation.

The other model considered in our analysis is called $2T$, which uses the generator $S_2={\rm Diag(-1,1,-1)}$ and the pair of values
$(p,m)=(3,3)$, which corresponds to the group $A_4$.
In this model   the mixing angles and the CP phase are related by the following relations:
\be
\sin^2\theta_{12} = \frac{1}{3 \cos^2\theta_{13}} 
\label{A4}
\ee
and
\be
\tan2\theta_{23} = \frac{1 - 2 s_{13}^2}{\cos \delta  s_{13}  
\sqrt{2 - 3 s_{13}^2}}. 
\label{eqqq}
\ee
The previous relations set important constraints on the reactor angle and the CP phase.
In fact, given the bounds on $\theta_{12}$, the reactor angle is restricted to be $\sin^2\theta_{13}\lesssim 0.025$, that is in a region where  
$\sin^2 \theta_{12}\sim 0.34$, still compatible with the range in eq.(\ref{bounds}). In addition, the allowed range for the atmospheric angle 
restricts $\delta$ to be below the maximal value. The situation is 
illustrated again in Fig.(\ref{figt1limits}), where we clearly see that the resulting parameter space in the $(\theta_{13},\delta)$-plane 
for this model is quite different 
from that of the $1T$ model.

\begin{figure}[h!]
\begin{center}
\includegraphics[scale=.44]{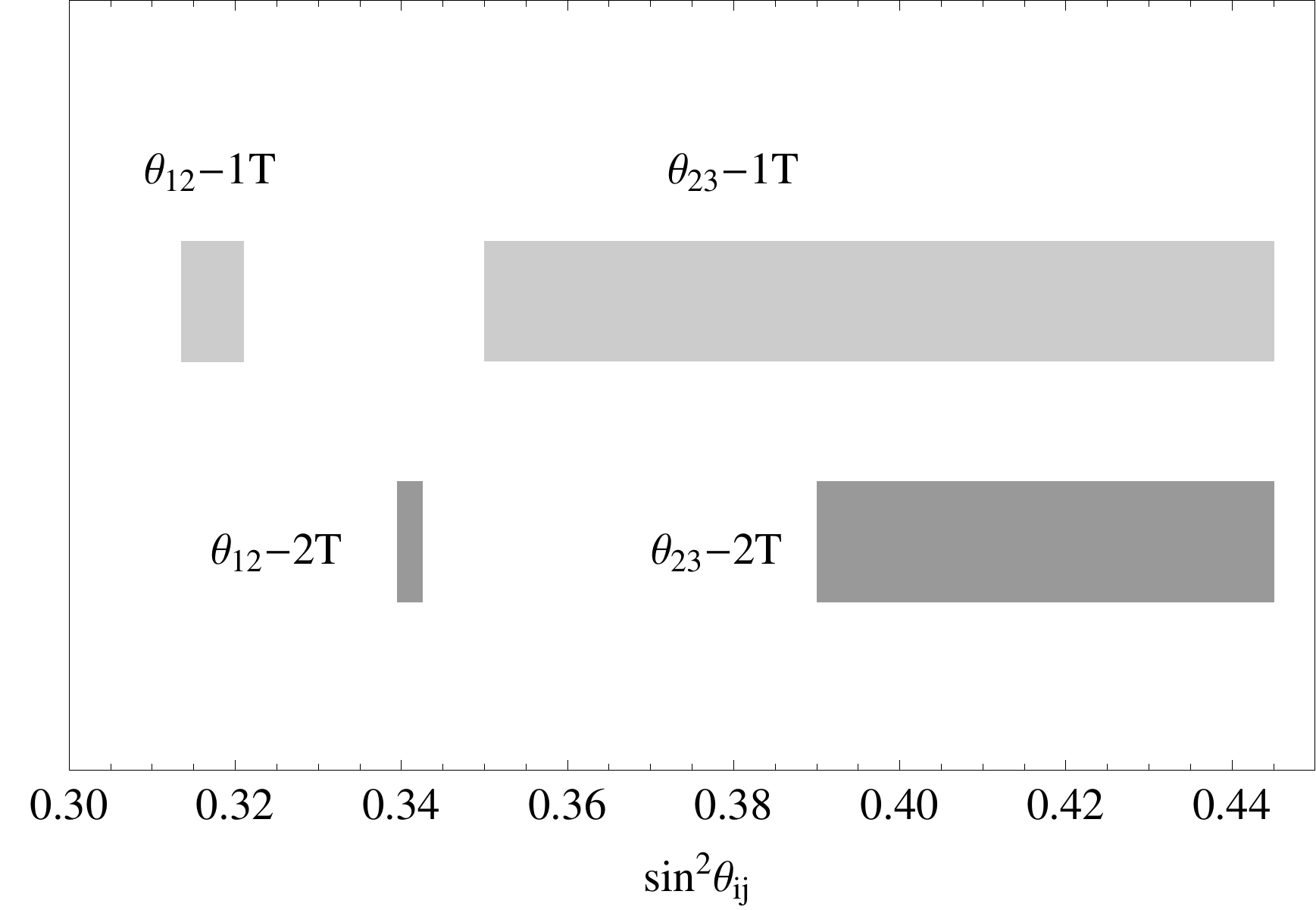} 
\caption{\label{figint}\it Allowed ranges for $\sin^2 \theta_{12}$ and $\sin^2 \theta_{23}$ for models $1T$ (light gray)
and $2T$ (dark gray).}
\end{center}
\end{figure}

The  allowed regions of the atmospheric and solar angles are instead summarized in Fig.(\ref{figint}).
As partially explained above, two very distinct intervals for the solar angles are implied by the two models, so a strong improvement in the 
measurement of the solar angle could be enough to distinguish among $1T$ and $2T$. On the other hand, there is a large overlap in the allowed
$\sin^2 \theta_{23}$, due to the still relatively large uncertainty affecting the determination of this angle.
In principle, a very precise measurement of $\sin^2\theta_{23}$ with central value well below $\sim 0.39$ can tell the two models but this possibility seems at 
the moment quite disfavored.

Equipped with the correlations of eqs.(\ref{S4a}-\ref{th23zzz}) and (\ref{A4}-\ref{eqqq}), we can now eliminate 
the dependence on $\theta_{12}$ and $\theta_{23}$ in the transition probabilities and 
get the expressions for the various $P_{\alpha\beta}$ as predicted by the models, $P^{1T,2T}_{\alpha\beta}$.
The resulting formulae are quite cumbersome, so we prefer not to show complicated analytical results that can hide the physical 
content of the probabilities. It is useful, instead, to study the differences:
\bea
\Delta P_{\alpha\beta} = |P^{1T}_{\alpha\beta}- P^{2T}_{\alpha\beta}|\,,\nn
\eea
which  give information on where to expect the largest differences among the two models.
To make life easier, we assume the same $\theta_{12}$ for the two models, which is a good approximation 
because the intervals of $\theta_{12}^{1T}$ and $\theta_{12}^{2T}$  
are contained in the 2$\sigma$ uncertainties quoted in eq.(\ref{bounds}), and to work in the intervals for $\theta_{13,23}$ where the models overlap; we take, however,
different CP phases, called $\delta_1$ if referred to the model $1T$ and $\delta_2$ if referred to $2T$.
We get:
\bea
\Delta P_{\mu e} &\sim& \frac{8}{3} \sqrt{2}\, r\,  \Delta\,  \sin \theta_{13}\, \sin \Delta\, \sin 
\left(\frac{\delta_{1}-\delta_{2}}{2}\right) \sin \left[\frac{1}{2} (2 \Delta +\delta_{1}+\delta_{2})\right]\nn  \\ &&\label{analitico} \\ \nn
\Delta P_{\mu \mu} &\sim& \frac{2}{3} \sqrt{2} \,r\,  \Delta\,  \sin \theta_{13} \,\sin 2\Delta\, \left(\cos\delta_{2}-\cos\delta_{1}\right)\,.
\eea
A common feature of the $\Delta P_{\alpha\beta}$'s is the leading dependence on $\theta_{13}$; given that 
$\sin\theta_{13}$ varies of about 17\% in the range quoted in (\ref{bounds}), we expect only a minor 
effects of $\theta_{13}$ in distinguishing the models, more pronounced for values close to the rightmost bounds of the regions 
in Fig.(\ref{figt1limits}).  
The dependence on the phases is, conversely, very significant.
In the case of the $\nu_\mu \to \nu_e$ appearance channel, $\Delta P_{\mu e}$ is sensibly different from zero for $\delta_1 - \delta_2 \sim \pi$;
this is not exactly the quantity separating the models under investigation since, as seen in Fig.(\ref{figt1limits}),
$\delta_1 - \delta_2 \in [0.6,2.4]$; however, this range of values guarantees that $\Delta P_{\mu e}\ne 0 $ and then that the 
two models can be, at least  in principle, distinguished.
For the  $\nu_\mu$ disappearance channel  we  observe that, for $\delta_1 \sim \pi/2 + \delta_2$ (whose approximate validity can 
be appreciated again from Fig.(\ref{figt1limits})), we obtain $\cos\delta_{2}-\cos\delta_{1}= \cos \delta_2 + \sin \delta_2 >0$, 
because $\delta_2 \lesssim \pi/4$; thus the phase dependence of  $\Delta P_{\mu \mu}$ 
results in the addition of two positive contributions and can be relevant in distinguishing the models.
 

As a final remark, we want to stress that our considerations remain valid for neutrino facilities where matter effects are small, 
which is the case of interest in this paper. 

\section{Models at long baseline neutrino experiments}
\label{sec3}
Having discussed the parameter space of the two models in an experimental-independent way, we now turn to the question 
on whether long baseline neutrino  facilities will be able to tell model $1T$ from model $2T$ based on the measurement 
of $P_{\alpha\beta}$'s and the CP-conjugate transitions. Our previous considerations on probabilities are drawn from analytical expressions in vacuum.
However, in studying the performance of a given experimental setup, one must take into account the experimental efficiencies 
to detect a given neutrino flavour. The expected number of events are computed according to:
\bea
N_\beta &=&N\,\int_{E_\nu} dE_\nu \,P_{\alpha \beta}(E_\nu)\,\sigma_\beta(E_\nu)\,\frac{d\phi_\alpha}{dE_\nu}(E_\nu) \,\varepsilon_\beta(E_\nu)\nn \\
&& \label{eq:asirate2}\\
\bar N_\beta &=&\bar N\,\int_{E_\nu} dE_\nu \,P_{\bar\alpha \bar\beta}(E_\nu)\,\sigma_{\bar\beta}(E_\nu)\,
\frac{d\phi_{\bar\alpha}}{dE_\nu}(E_\nu) \,\varepsilon_{\bar \beta}(E_\nu) \nn\,,
\eea
in which $\sigma_{\beta (\bar \beta)}$ is the cross section for producing the lepton $\beta(\bar \beta)$,  
$\varepsilon_{\beta(\bar \beta)}$ the detector efficiency to reveal that lepton, $\phi_{\alpha(\bar \alpha)}$ the initial neutrino flux at 
the source and $N$, $\bar N$ are normalization factors containing the detector mass and the number of years of data taking.  These events can also be grouped in neutrino energy bins, thus taking full advantage of different spectral information
of $P_{\alpha \beta}$ or $P_{\bar\alpha \bar\beta}$.
Since the probabilities we are interested in are $P_{\mu e}$ and $P_{\mu\mu}$, 
$\beta$ can be an electron or a muon  (and their antiparticles).

In order to assess the capabilities of a given facility to tell $1T$ from $2T$ we adopt the following strategy:
\begin{itemize}
\item for any pair of the mixing parameters $(\bar \theta_{13},\bar \delta)$ in the regions allowed for the model $2T$ 
we compute the expected number of events $N^{2T}_{\alpha,i}(\bar \theta_{13},\bar \delta)$ for a given final flavour $\alpha$ and 
neutrino energy bin $i$ ($\theta_{12}$ and $\theta_{23}$ are then 
determined from eqs.(\ref{A4})-(\ref{eqqq})); 
\item we then compare $N^{2T}_{\alpha,i}(\bar \theta_{13},\bar \delta)$ to $N^{1T}_{\alpha,i}(\theta_{13},\delta)$, where now 
the mixing parameters are those of the competing model $1T$. In this procedure, we are implicitly  assuming that the model $2T$ 
and the pair $(\bar \theta_{13},\bar \delta)$ are the one chosen by Nature;
\item in the next step, we minimize the following $\chi^2$ variable over $\theta_{13}$ and $\delta$ in the $1T$ allowed parameter space
\cite{Meloni:2012fq}:
\bea
\label{chi2}
\chi^2 =
\Sigma_{\alpha,i} \frac{\left[N^{1T}_{\alpha,i}(\theta_{13}, \delta)-N^{2T}_{\alpha,i}(\bar \theta_{13},\bar \delta)\right]^2}{\sigma^2_{\alpha,i}}\,,
\eea
where the uncertainty is given by:
\bea
\sigma^2_{\alpha,i}=N^{2T}_{\alpha,i}(\bar \theta_{13},\bar \delta)+B_{\alpha,i}+(n_\alpha\,
N^{2T}_{\alpha,i}(\bar \theta_{13},\bar \delta))^2+(b_\alpha\,B_{\alpha,i})^2\,,
\eea
in which $B_{\alpha,i}$ is the background associated to $N^{2T}_{\alpha,i}(\bar \theta_{13},\bar \delta)$, $n_{\alpha}$  
the overall systematic error related to the determination of $N_{\alpha(\beta),i}$ and  $b_{\alpha}$ that of $B_{\alpha,i}$.
For the sake of simplicity,  $n_{\alpha}$ and $b_{\alpha}$ are constant in the whole energy range;
\item if the obtained minimum is larger 
than some reference $\chi^2$ value, $\chi^2_{min}\ge\chi^2_{cut}$, then in the point $(\bar \theta_{13},\bar \delta)$ the two models
can be distinguished at a given confidence level. The ensemble of such points identifies the wanted regions.
\end{itemize}
Obviously, the procedure can also be applied in the reverse order, that is considering $1T$ as the {\it true} model and 
finding a minimum of the $\chi^2$ function in the $2T$ parameter space. The results will 
then be  presented in the $1T$ $( \theta_{13},  \delta)$-plane.

In the following numerical simulations, we will proof our strategy at three different experimental setups: NO$\nu$A, T2K and T2HK.
All events rates are computed using exact numerical probabilities in matter.

\subsection{Results from NO$\nu$A$\oplus$T2K and T2HK}
\begin{figure}[h!]
\begin{center}
\includegraphics[scale=.44]{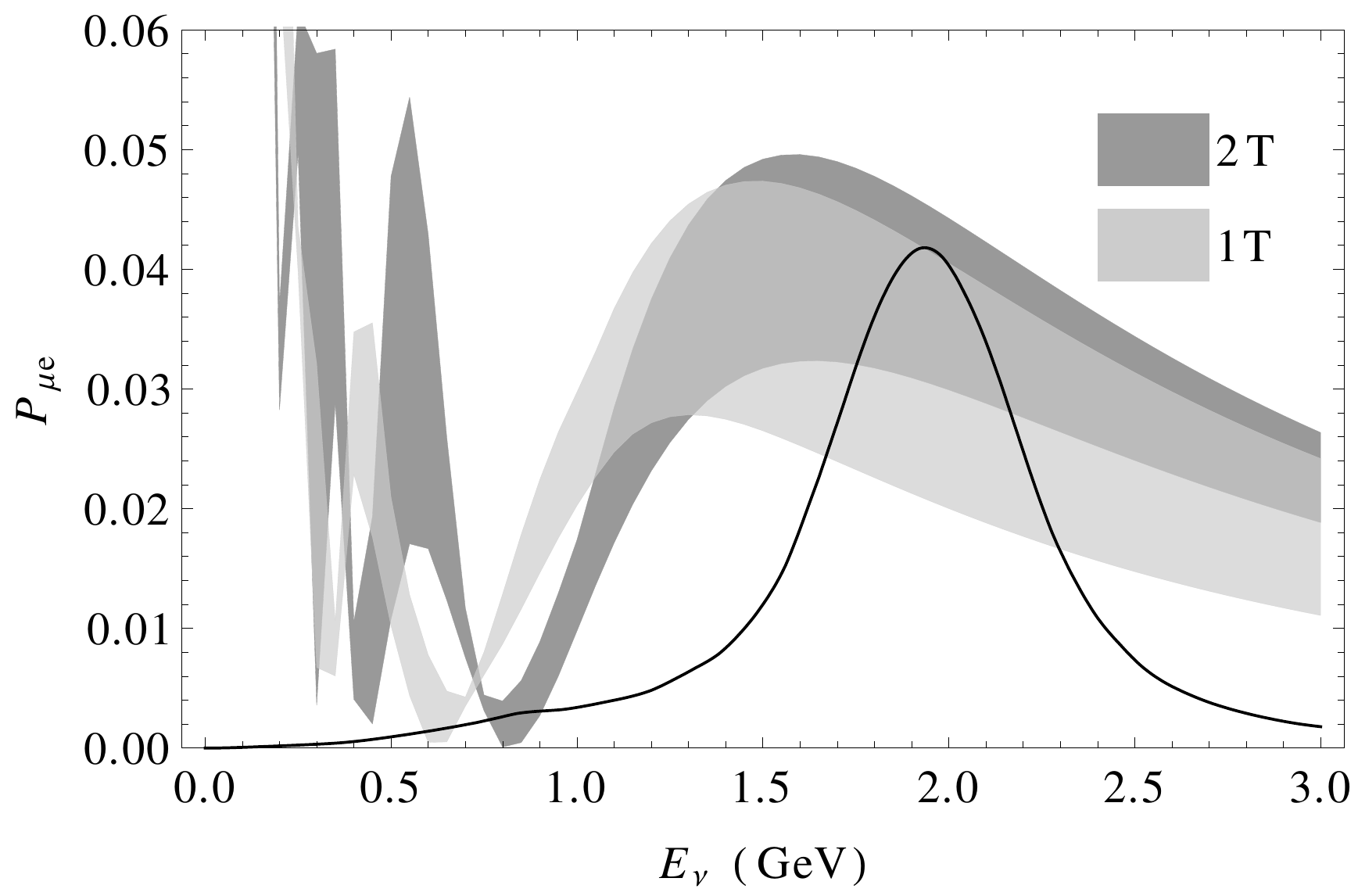}\includegraphics[scale=.44]{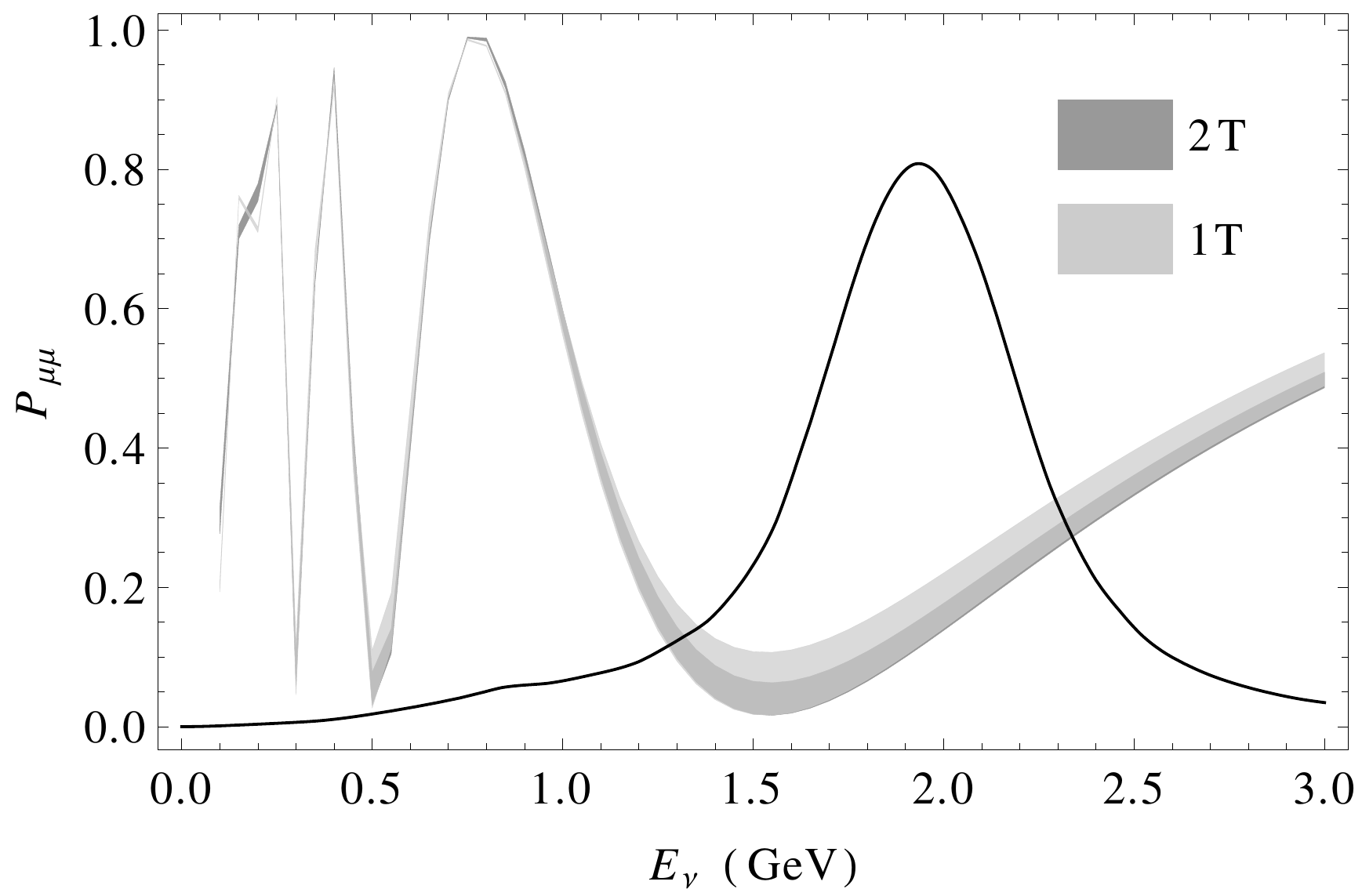}
\caption{\label{figprob}\it Range of values of $P^{1T,2T}_{\mu e}$ (left panel) and $P^{1T,2T}_{\mu \mu}$ (right panel)
as a function of the neutrino energy, for the NO$\nu$A setup. The solid line is a down-scaled version of the $\nu_\mu$ NO$\nu$A flux, in arbitrary units.}
\end{center}
\end{figure}

\begin{figure}[h!]
\begin{center}
\includegraphics[scale=.5]{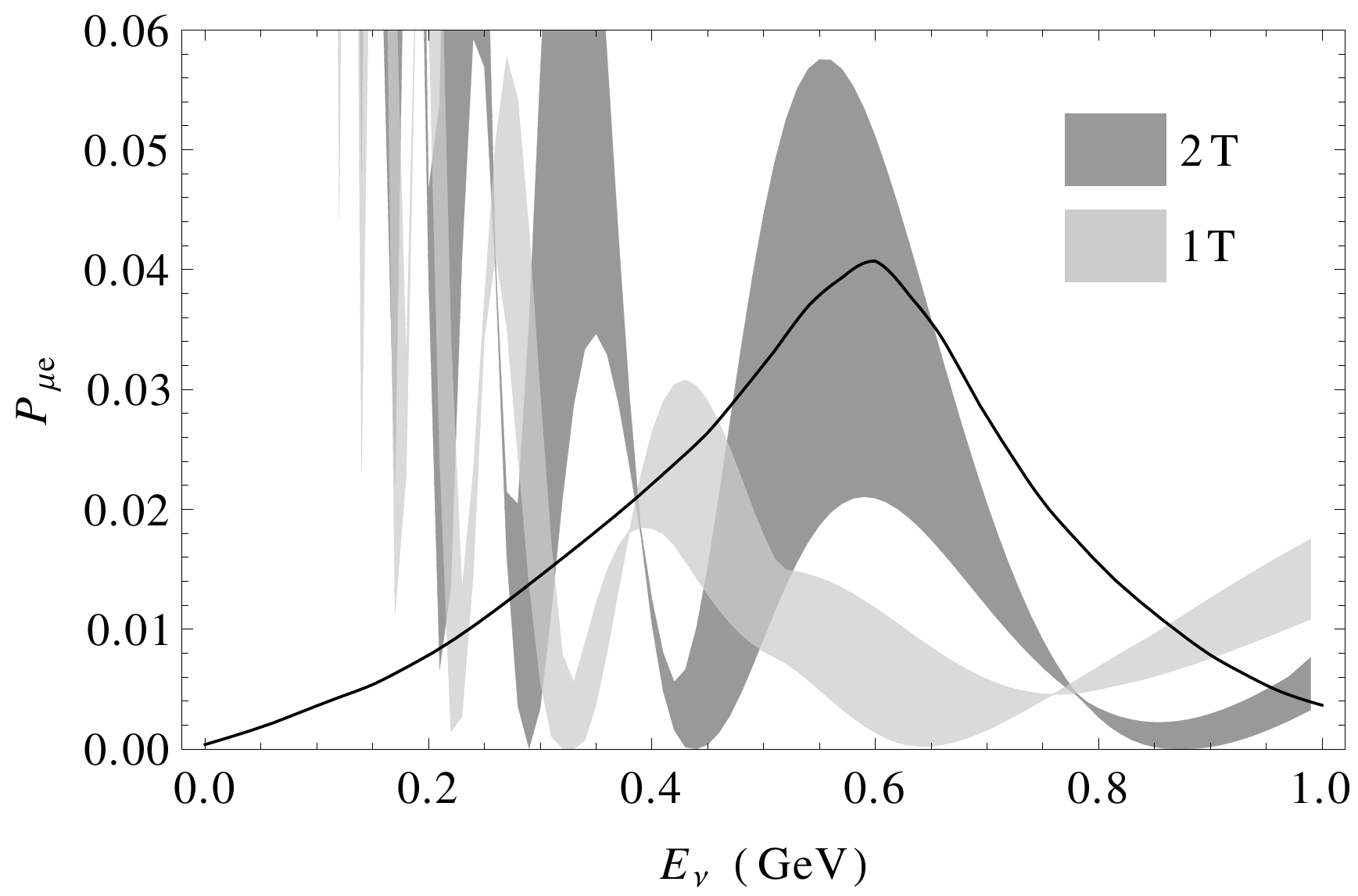} 
\caption{\label{figprob2}\it Range of values of $P^{1T,2T}_{\mu e}$ 
as a function of the neutrino energy, for the T2K setup. The solid line is a down-scaled version of the $\nu_\mu$ T2K flux, in arbitrary units.}
\end{center}
\end{figure}

In this section, we briefly consider the experimental setups used in our numerical simulations.
\begin{itemize}
 \item the {\bf NO$\nu$A} detector \cite{Patterson:2012zs}
is a 14 $\mathrm{kt}$ totally active scintillator detector (TASD), located at a distance 
of 810 $\mathrm{km}$ from Fermilab, with an off-axis angle of  $0.8^\circ$  
from the NuMI beam. 
In the appearance mode \cite{Ayres:2004js}, the main
backgrounds are due to the intrinsic beam $\nu_e/\bar{\nu}_e$,
mis-identified muons  and single $\pi^0$ events
from neutral current interactions. In the disappearance mode \cite{tyang}, 
we have to consider  wrong-sign muon from
$\bar{\nu}_\mu~(\nu_\mu)$ contamination in $\nu_\mu (\bar{\nu}_\mu)$
beam and neutral current events.
Due to the relatively large $\theta_{13}$, 
the collaboration has relaxed the cuts for the event selection
criteria,  allowing for more signal events  along with more background
events \cite{Agarwalla:2012bv}. Our simulation is mainly based on the files provided by the GLoBES software \cite{globes,Huber:2009cw},
with migration matrices from \cite{Akiri:2011dv} and kindly provided by one of the authors of \cite{Coloma:2012ji}.
In this way, the signal and backgrounds events  released by the NO$\nu$A Collaboration are 
reproduced \cite{Patterson:2012zs}. For the sake of simplicity, we take all systematics effects 
at the level of 5\%, that is $n_\alpha=b_\alpha=0.05$ for  $\alpha=e^{-},e^{+},\mu^{-},\mu^{+}$.
\end{itemize}
For fixed $\alpha$ and $\beta$, the energy dependence of the probabilities
$P^{1T}_{\alpha\beta}$ and $P^{2T}_{\alpha\beta}$ is quite different so, 
at least in principle, the facility could  be efficient in discriminating the two models.
In Fig.(\ref{figprob})
we show both $P^{1T}_{\alpha\beta}$ (in light gray) and $P^{2T}_{\alpha\beta}$ (in dark gray) as a function of the neutrino energy,
obtained varying all the mixing parameters in the respective allowed ranges,  with $(\alpha,\beta)=(\mu,e)$ 
in the left panel and $(\alpha,\beta)=(\mu,\mu)$ in the right one.
The solid line is a down-scaled version of the $\nu_\mu$ NO$\nu$A flux.
Beside the large fluctuations of the probabilities below $E_\nu \lesssim 1$ GeV, which are less important due to the smallness of the 
neutrino flux, in both cases $P_{\mu e}$ and $P_{\mu \mu}$ show a different behavior close to the maximum of the flux, that is in an energy 
region where NO$\nu$A will collect the bulk of the events. In particular, in the appearance channel the spread we observe for the $1T$ model is 
mainly a consequence of a larger uncertainty on $\theta_{23}$, which reaches values smaller than $\sin^2 \theta_{23}\sim 0.39$ and then
makes $P^{1T}_{\mu e} \lesssim P^{2T}_{\mu e}$ close to the pick. A smaller atmospheric angle also means a larger $\nu_\mu$ disappearance,
so in the left plot we have $P^{1T}_{\mu e} \gtrsim P^{2T}_{\mu e}$ for energies above 1 GeV.
%
%
\begin{itemize}
 \item for the {\bf T2K} we consider the Super-Kamiokande water Cerenkov detector
 of fiducial mass of 22.5 $\mathrm{kt}$, placed at a distance of 295 $\mathrm{km}$ from
the source beam from J-PARC, at an off-axis angle of $2.5^\circ$. 
Our numerical simulation have been performed based to the information provided in the corresponding GLoBES 
files, described in 
\cite{Huber:2009cw,Fechner:2006koa}, to which we refer for details.
\end{itemize}
The appearance channels in T2K show an even increased capability to distinguish among  $P^{1T}_{\mu e}$ and $P^{2T}_{\mu e}$: in fact, 
$P^{1T}_{\mu e}$ is generally smaller than $P^{2T}_{\mu e}$ for energies at and below the 
maximum of the T2K flux, thus making the prediction of the two models significantly different, see Fig.(\ref{figprob2}).
On the other hand, for the disappearance channel we do not observe such a huge difference and we do not present 
the corresponding plot. 
\begin{itemize}
 \item for the {\bf T2HK} setup we follow the proposal  and the
  Letter of Intent presented in ~\cite{Abe:2011ts}, with  a WC detector with a fiducial mass of 560~kton, placed at a
  distance of 295~km from the source. We assume again $n_\alpha = b_\alpha = 0.05$.
\end{itemize}

\begin{figure}[h!]
\begin{center}
   \includegraphics[scale=.44]{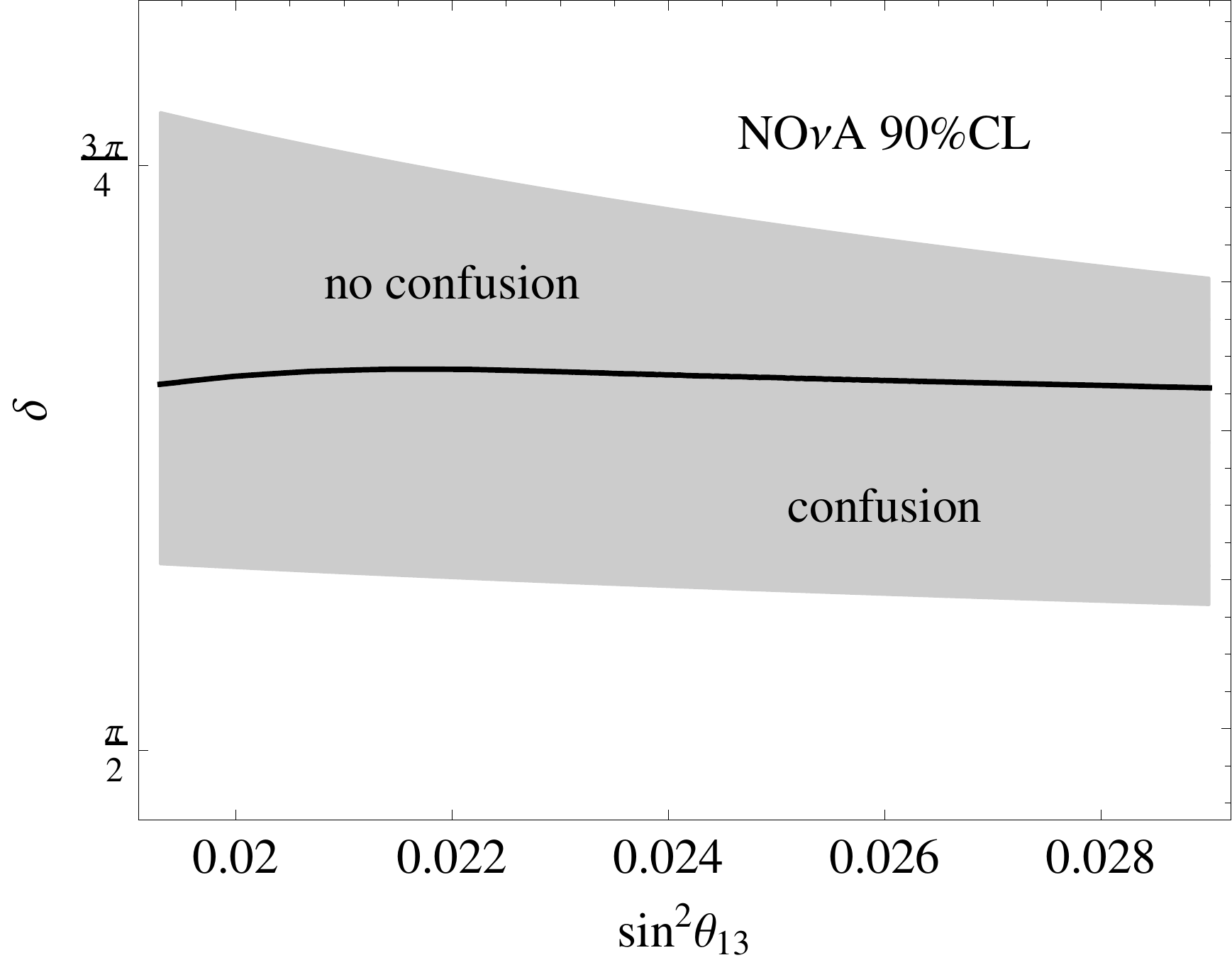} \qquad \includegraphics[scale=.44]{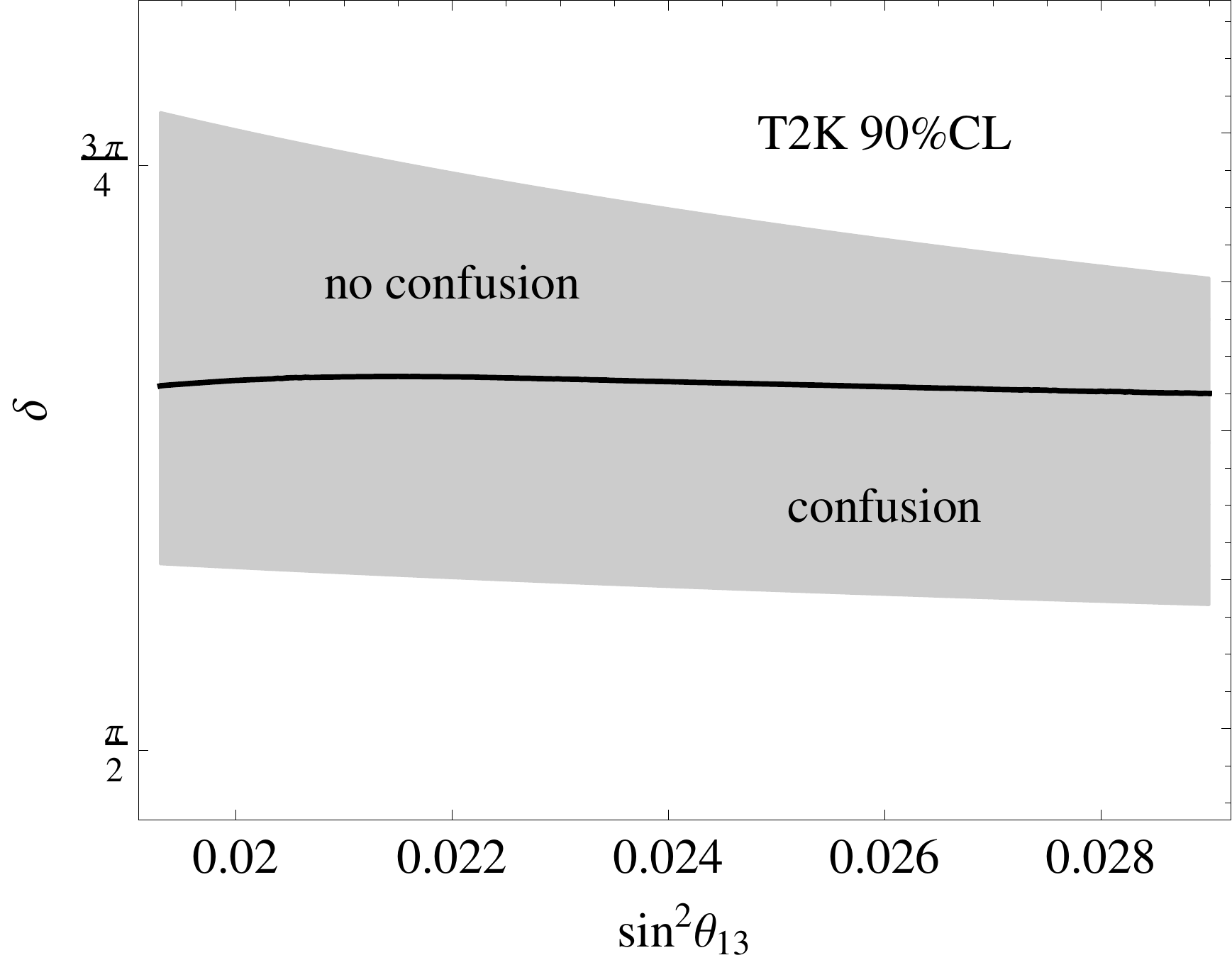}
\caption{\label{figcomp}\it Regions in the $1T$ parameter space where the $1T$ and $2T$ models can be distinguished at 90\% confidence level, using the
appearance and disappearance channels. Left plot: for the NO$\nu$A setup. Right plot:  for the T2K setup.}
\end{center}
\end{figure}

It is clear that NO$\nu$A and T2K, taken individually, have the potential to make some sort of discrimination among the $1T$ and 
$2T$ models which, however, strongly depends on the assumed values of $n_\alpha$, so 
NO$\nu$A and T2K can say something relevant only in a limited portion of the parameter space. In particular, we have found that no 
distinction is possible if we assume that $2T$ is the {\it correct} model, for any value of $n_\alpha$. 
On the other hand, under the assumption that $1T$ 
gives the values of the mixing parameters chosen by Nature and $n_\alpha=0.05$, a limited discrimination is possible for those points in the 
$(\theta_{13},\delta)$-plane with the largest possible values of the CP-phase, in agreement with our discussion below eq.(\ref{analitico}).
This can be seen in Fig.(\ref{figcomp}) where we show the results of our computation in the $(\theta_{13},\delta)$-plane allowed for the $1T$ model, 
in the case on NO$\nu$A (left plot) and T2K (right plot) experimental setups. 
In both plots, the points above the solid lines, $\delta \gtrsim 2.06$, identify the region where the two models can be distinguished at the 90\% 
of confidence level, using both appearance and disappearance channels. 
As expected, the capability of the considered facilities to distinguish the two models is almost independent on the value of $\theta_{13}$,
as emphasized in the previous section. For values of $n_\alpha$ as large as 10\% no distinction is possible.
The sensitivities are the results of a strong synergy among the appearance and disappearance channels; in fact, 
we have observed that:
\begin{itemize}
\item the appearance channel alone  cannot give any useful information, since the sensitivity lines lie above the maximum values of 
$\delta$ in the $1T$ parameter space;
\item  the $\nu_\mu \to \nu_\mu$ transition alone does not allow any discrimination among $1T$ and $2T$, given the mild 
dependence in $P_{\mu\mu}$ on $\theta_{13}$ and $\delta$, see eq.(\ref{eq:Pmumu}). However, when used in combination with the $\nu_\mu \to \nu_e$  
channel, the 
disappearance transition sorts some effects, due to the ability of measuring $\theta_{23}$ whose allowed ranges are slightly  different
in the two models. Although we fixed the solar and mass differences to their best fit values quoted in \cite{Fogli:2012ua}, the inclusion 
of the uncertainty on $\Delta m^2_{31}$ (and, to a less extent, the one from  $\Delta m^2_{21}$) does not change 
appreciably the regions where confusion is avoided, mainly due to the relatively small error on 
$\Delta m^2_{31}$ at the 2$\sigma$ level, around 4-5\% for both NO$\nu$A and T2K facilities.
\end{itemize}

\begin{figure}[h!]
\begin{center}
\includegraphics[scale=.44]{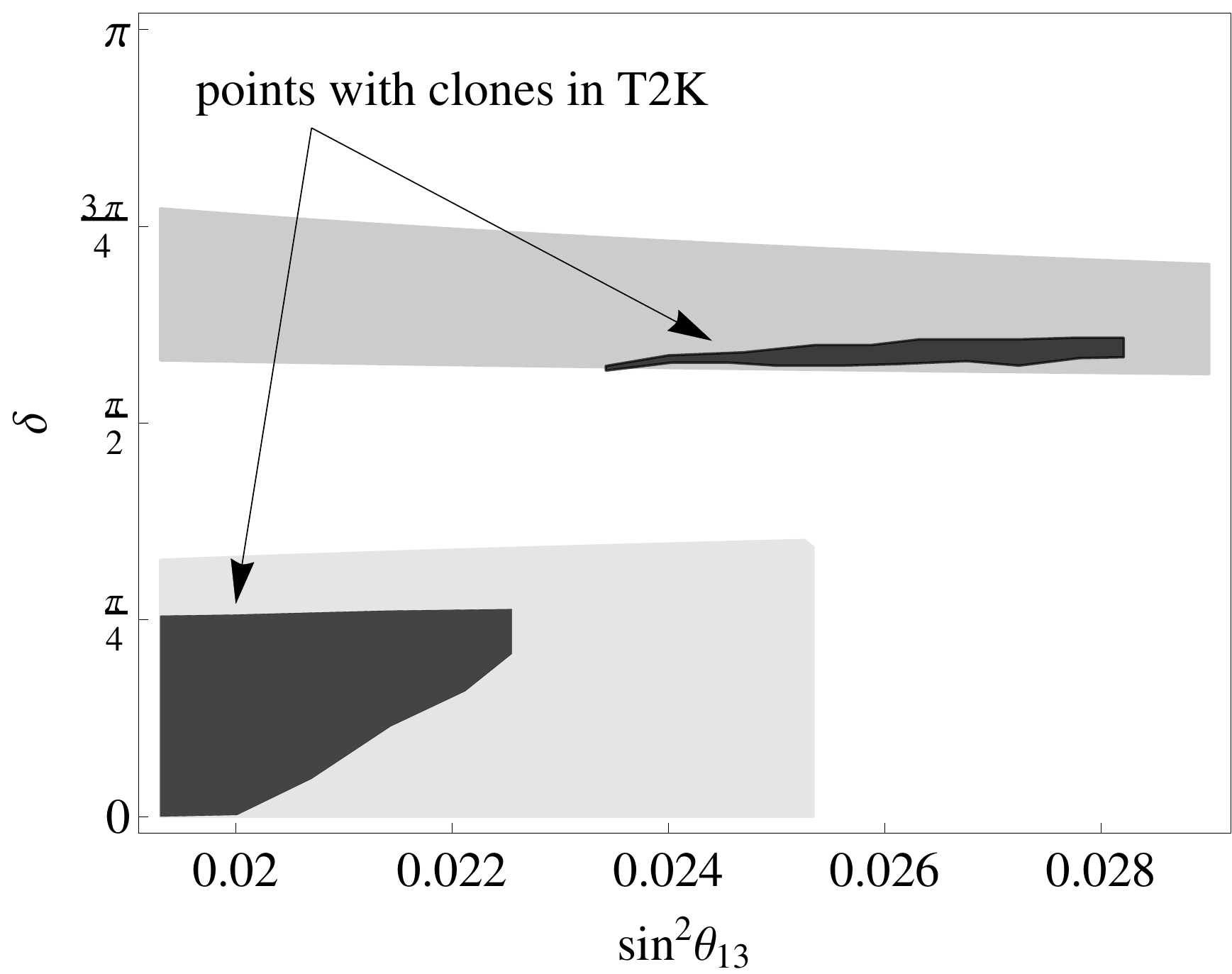} 
\caption{\label{figclones}\it Values of $\theta_{13}$ and $\delta$ in the $1T$ and $2T$ parameter spaces where eq.(\ref{clo}) (and the one obtained with 
the replacement $1T \leftrightarrow 2T$) is solved. 
The computation has been performed for the T2K setup and considering total rates only.}
\end{center}
\end{figure}

A different situation arises if we combine the simulated data 
from both experiments.  The most interesting feature is that a  (reduced) region in the $2T$ parameter space appears where 
the two models can be distinguished.
\begin{figure}[h!]
\begin{center}
\includegraphics[scale=.44]{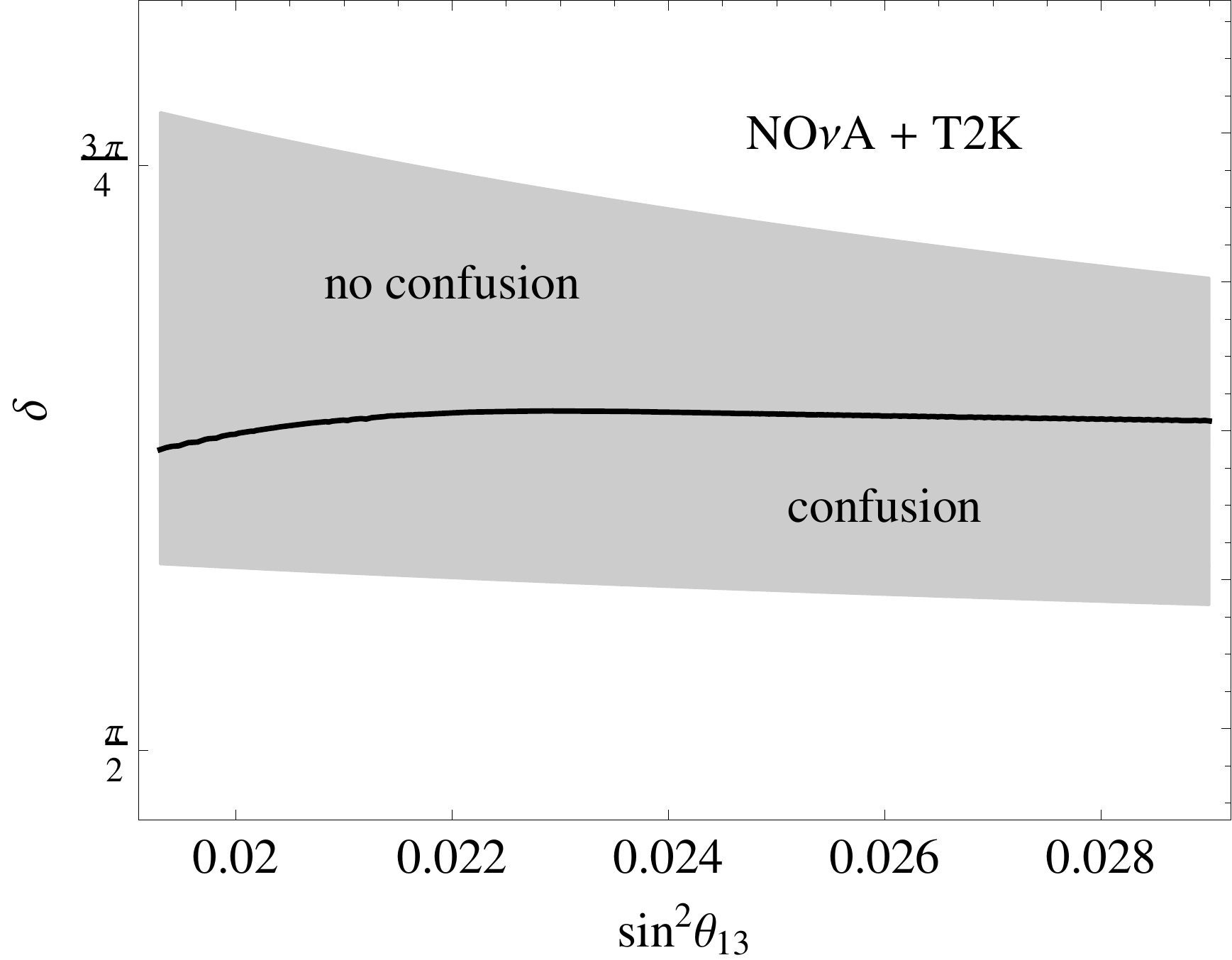} \qquad \includegraphics[scale=.44]{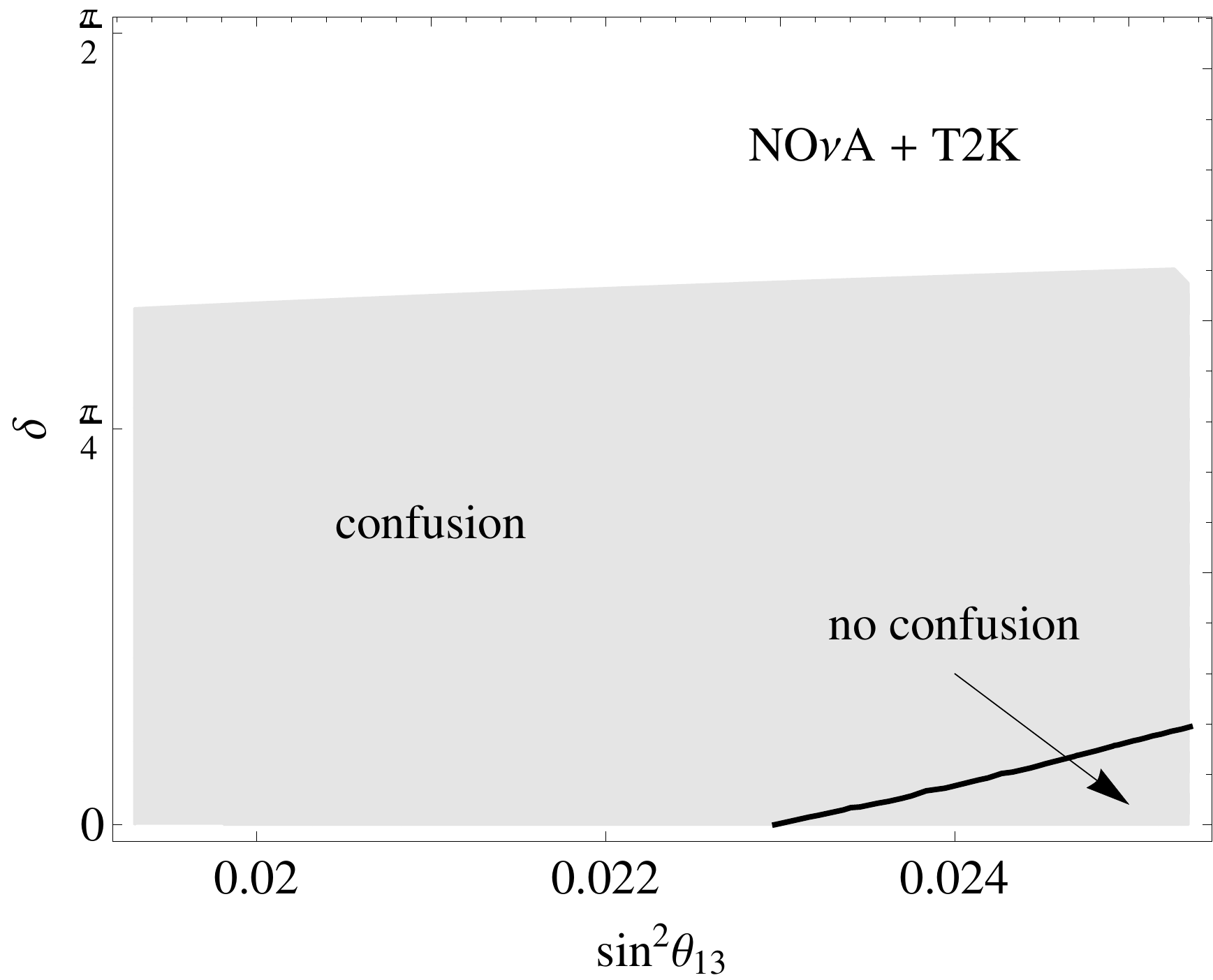}
\caption{\label{figcomp2}\it Regions in the $1T$ parameter space (left panel) and $2T$ parameter space (right panel)
where the two models under investigation  can be distinguished at 90\% confidence level, combining the results from both NO$\nu$A and T2K.}
\end{center}
\end{figure}
It involves values of $\delta$ no larger than 0.2, and only for values of the reactor angle close to its upper bound.
In the $1T$ parameter space, we observe only a modest improvement with respect to the case of Fig.\ref{figcomp}, due to the 
fact that the $\chi^2$ functions of the two setups are very similar in the portion of the parameter space considered, so that
no  powerful synergy is at work when combining the data.
The different sensitivities observed in the $1T$ and $2T$ $(\theta_{13},\delta)$-plane are easily understood in terms 
of {\it intrinsic clones} \cite{Donini:2003vz}, that is in terms of points in the parameter space with the same number of expected events. 
Consider first the $2T$ model;  the minimum of the $\chi^2$ in eq.(\ref{chi2}) is expected to appear close to the points where the system of equations:
\bea
N^{2T}_{\mu,i}(\bar \theta_{13},\bar \delta) &=& N^{1T}_{\mu,i}(\theta_{13},\delta)  \nonumber \\ && \label{clo}\\
N^{2T}_{e,i}(\bar \theta_{13},\bar \delta) &=& N^{1T}_{e,i}(\theta_{13},\delta)  \nonumber
\eea
has a solution for $(\theta_{13},\delta)\ne (\bar \theta_{13},\bar \delta)$. 
A numerical scan of the pairs $(\bar \theta_{13},\bar \delta)$ in the $2T$ parameter space, performed using the total event rates, 
has shown that many points with small $\bar \theta_{13}$ have a mirror
in the $1T$ plane at values close to the smaller allowed $\delta$  and large $\theta_{13}$. 
Such  $(\bar \theta_{13},\bar \delta)$ pairs are then not good to perform a discrimination.
Changing $1T \leftrightarrow 2T$ into eq.(\ref{clo}) produces very similar results, in the sense that the region
that was before  the {\it mirror region} is now made of the $(\bar \theta_{13},\bar \delta)$ pairs in the $1T$ parameter
space where discrimination is not possible (as they have clones located in the $2T$ space at small $\theta_{13}$). For the T2K setup, these  regions  (black areas) are presented in Fig.(\ref{figclones}).
Taking into account that the solution of eq.(\ref{clo}) only give an indicative position of the clone points and that 
NO$\nu$A has roughly the same ($L/E_\nu$) as T2K, Fig.(\ref{figclones})  
shows many of the features of the allowed regions presented in Fig.\ref{figcomp2}: the distinction of the models in 
the $1T$ space happens at the largest possible values of $\delta$ and that in the $2T$  can happen only at large $\theta_{13}$.

\begin{figure}[h!]
\begin{center}
\includegraphics[scale=.44]{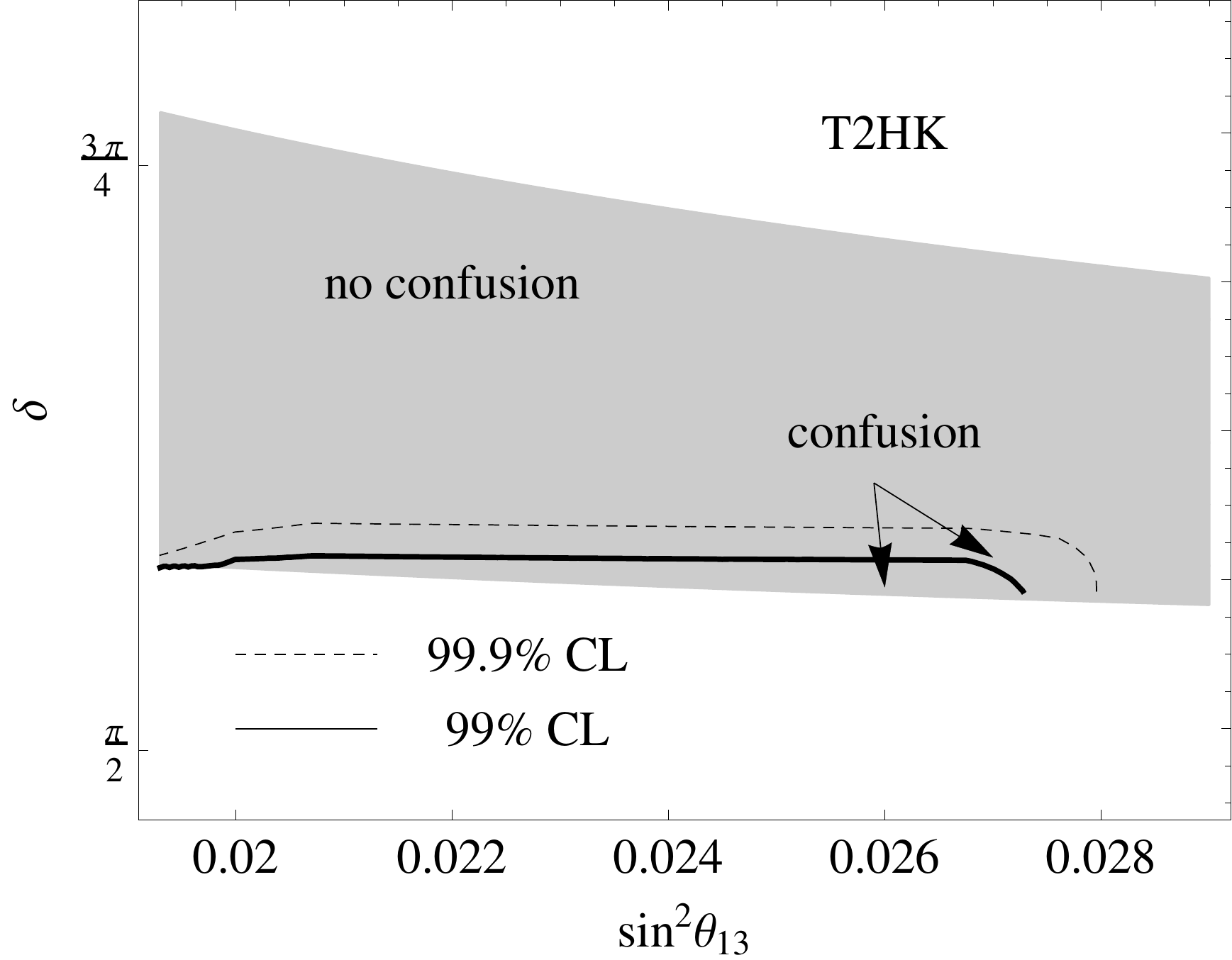} \qquad \includegraphics[scale=.44]{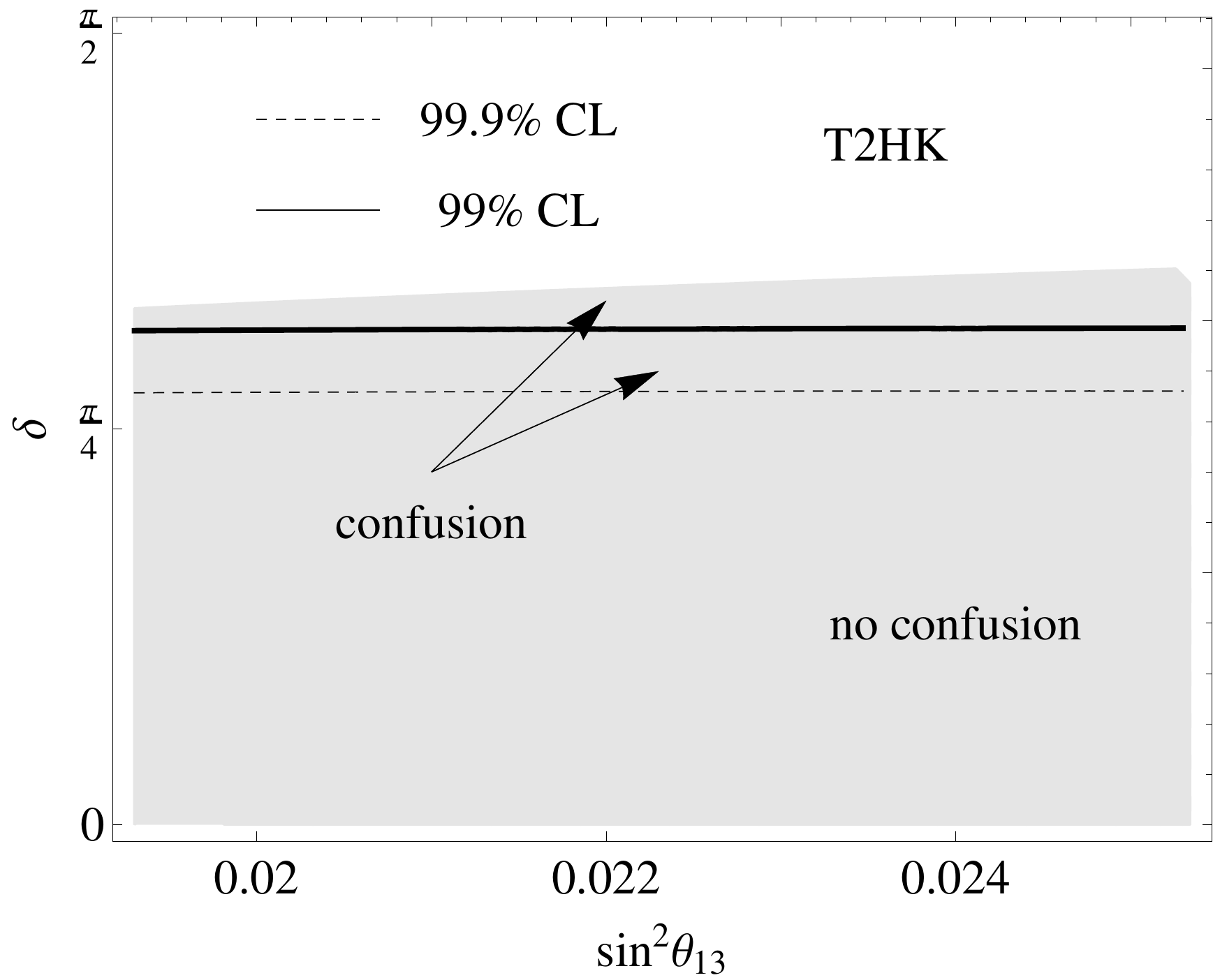}
\caption{\label{plotst2hk}\it Regions in the $1T$ parameter space (left panel) and $2T$ parameter space (right panel)
where the two models under investigation  can be distinguished at 99\% confidence level (solid line) and 
99.9\% confidence level (dashed line), in the case of the T2HK experimental setup.}
\end{center}
\end{figure}

For the T2HK setup, we get a much better capability of distinguishing the models, Fig.(\ref{plotst2hk}); 
in fact, in both $1T$ and $2T$ parameter spaces the 
regions where confusion is possible (at 99\% and 99.9\% CL) are confined into thin stripes close to the lower ($1T$) and upper ($2T$) bounds, 
thus making this facility quite appropriate for model selection. The good performance with respect to the T2K setup has to be 
ascribed to the interplay between a larger detector mass and the use of the antineutrino modes. In particular, we 
have verified that the inclusion of the antineutrino mode into the analysis is crucial to get the sensitivities shown
in Fig.(\ref{plotst2hk}) which, otherwise, would be a rescaled version of the T2K results shown in the right panel of 
Fig.(\ref{figcomp}) in the $1T$ parameter space, and a reduced sensitivity (for small $\delta$ 
and large $\theta_{13}$) in the $2T$ parameter space.

A summary of the previous considerations is presented in Tab.\ref{tab2} where, for each of the facilities and combination analyzed above, 
we reported our estimates of the range of values of the CP phase where distinction is possible among the $1T$ and $2T$ models. 
These ranges are obviously modulated by $\theta_{13}$ (and in the table we use "{\it upper bound}" to indicate 
the upper border of the $1T$ allowed parameter space), so they represent indicative intervals.

\begin{table}[h!]
\centering
\begin{tabular}{ |l||c|c|c|c|} 
\hline
\hline
\multicolumn{5}{|c|}{\it Approximate ranges in $\delta$}\\
\hline \hline
\multicolumn{1}{|c||}{ }
&\multicolumn{1}{c|}{{ \nova}}
&\multicolumn{1}{c|}{{T2K}}
&\multicolumn{1}{c|}{{\nova + T2K}}
&\multicolumn{1}{c|}{{T2HK (99\% CL)}} \\
\hline
        $1T$
         &[2.06,upper bound] &[2.06,upper bound]
         &[2,upper bound] &[1.83,upper bound]  \\
         \hline
         $2T$
         &- &-
         &[0,0.1] for large $\theta_{13}$ &[0,1]\\
\hline
\hline
\end{tabular}
\caption{\it Estimates of range of values of $\delta$ where distinction is possible among the $1T$ and $2T$ models for 
the facilities analyzed in this paper. "Upper bound" refers 
the upped border of the allowed region for the $1T$ model. Dashes indicate that no discrimination is possible.}
\label{tab2}
\end{table}


\section{Conclusions}
\label{conc} 
Starting from two different neutrino mixing sum rules  we have studied if, and to which extent, NO$\nu$A, T2K and T2HK 
are able to falsify one of them in favor of the other. This is due to the fact that 
the two sum rules identify different set of values of the neutrino mixing parameters, namely 
different regions in the CP phase $\delta$ and $\theta_{12}$ and partially overlapping regions for $\theta_{13}$ and $\theta_{23}$,
all of them compatible with the experimental values at 2$\sigma$. Analytical considerations on the $\nu_\mu \to \nu_e$ and 
$\nu_\mu \to \nu_\mu$ transition probabilities revealed that distinguishing the two type of correlations is possible 
for large differences among the true values (chosen in one parameter space) and the fitted values (in the competing parameter space) 
of $\delta$. Our numerical simulations have shown that this is indeed the case; in particular,  NO$\nu$A and  T2K taken alone have the capabilities 
to tell the $1T$ model from the $2T$ model at 90\% of confidence level, reducing the portion in the $(\theta_{13},\delta)$-plane of the 
$1T$ model where confusion is possible. In the $2T$ parameter space we revealed a much worse performance, unless the combination of NO$\nu$A +  T2K data is 
taken into account, and only in a very limited region at large $\theta_{13}$ and small $\delta$. On the other hand, the 
T2HK experimental facility, taking full advantage of a larger detector mass and of the use of the $\bar \nu_\mu$ flux compared to 
the T2K setup, has a much better performance in terms of model selection
in both parameter spaces, leaving aside only a small portion of values of $\delta$ where confusion is still possible. 
These small regions 
disappear if we consider the setup of the {\sf NF10}, thus making this facility useful to perform a selection of sum rules 
modified by the inclusion of various type of next-to-leading order effects.

\subsection*{Acknowledgments}
We are strongly indebted with Pilar Coloma, for kindly providing the modified GLoBES files to simulate 
NO$\nu$A and {\sf NF10}. We also want to acknowledge Enrique Fernandez Martinez for useful suggestions 
on the first version of the paper.
We acknowledge MIUR (Italy) for financial support under the program ”Futuro in Ricerca 2010 (RBFR10O36O)”
and CERN, where this work has been completed.

\end{document}